\documentstyle[twocolumn,prl,aps,epsfig,amssymb]{revtex}

\begin{document}
\draft
\title{Can MSSM with light sbottom and light gluino survive $Z$-peak constraints ? }

\author{Junjie Cao $^{1,2}$, Zhaohua Xiong $^{1,3}$, Jin Min Yang $^3$}
 
\address{$^1$ CCAST (World  Laboratory), P.O.Box 8730, Beijing 100080, China}
\address{$^2$ Physics Department, Henan Normal University, Henan 453002, China}
\address{$^3$ Institute of Theoretical Physics, Academia Sinica, Beijing 100080, China}
\date{\today}
\maketitle
 
\begin{abstract}

In the framework of minimal supersymmetric model we examine
the $Z $-peak constraints on the scenario of one light sbottom 
($2\sim 5.5$ GeV) and light gluino ($12\sim 16$ GeV), which has been 
successfully used to explain the excess of bottom quark production 
in hadron collision.  Such a scenario is found to be severely constrained 
by LEP $Z$-peak observables, especially by $R_b$, due to the large effect 
of gluino-sbottom loops. To account for the $R_b$ data in this scenario, 
the other mass eigenstate of sbottom, i.e., the heavier one,  must be 
lighter than $125$ ($195$) GeV at $2\sigma$ ($3 \sigma $) level, which
should have been produced in association with the lighter one at LEP II
and will probobaly be within the reach of Tevatron Run 2. 
    
\end{abstract}

\noindent{\bf Introduction}~~
Although the standard model (SM) has been successful phenomenologically, it is 
generally believed to be an effective theory valid at the  electroweak scale
and some new physics must exist beyond the SM. This belief was seemingly 
corroborated by some experiments, such as the recent measurement of muon 
$g-2$ \cite{E821} and the evidence of neutrino oscillations~\cite{superK}. 
Among various speculations of new physics theories, the minimal supersymmetric 
model (MSSM) is arguably a promising candidate and has been intensively 
studied in the past decades.

The non-observation of any sparticles from direct experimental searches 
suggested heavy masses for sparticle spectrum. However, there have been 
a lot of analysis \cite{lightbg} which argue that a very light sbottom 
and light gluino (with mass of a few GeV)   may have escaped from the 
direct experimental searches.  It is intriguing that a light sbottom 
may require a light gluino, as analyzed in the last reference in 
\cite{lightbg}.  A recent analysis \cite{doelectr} 
showed that a light sbottom ($\tilde{b}_1$) with mass comparable with 
bottom quark is still allowed by electroweak precision  data if its 
coupling to $Z$ boson is small enough.  A study by Berger {\em et~al}
\cite{Berger01} found that the scenario of MSSM with one light sbottom 
($2 \sim 5.5$ GeV) and a light gluino ($12\sim 16$ GeV) can successfully 
provide an explanation for the long-standing puzzle that the measured 
cross section of bottom quark production at hadron collider exceeds 
the QCD prediction by about a factor of 2 \cite{CDF93}. They also argued 
that such a scenario is consistent with all experimental constraints 
on the masses and couplings of sparticles. 

We note that the previous examinations \cite{doelectr} on 
$Z$-peak constraints focus on the direct  production of a light sbottom
followed by its decay similar to the bottom quark. Then by fine-tuning 
the mixing of left- and right-handed sbottoms, the coupling of $Z$-boson 
to the lighter mass eigenstate of sbottom ($\tilde{b}_1$)  can be 
sufficiently small so as to avoid the  $Z$-peak constraints. It is 
noticeable that when sbottom  $\tilde{b}_1$ and gluino are both light, 
as was used to explain the excess of bottom quark production in hadron 
collision \cite{Berger01}, gluino-sbottom loops may cause large effects 
in $Zb\bar b$ coupling \footnote{Previous calculations of SUSY loop 
effects on $Zb\bar b$ coupling focused on rather heavy squarks and 
gluino and thus obtained very small effects\cite{Djouadi91}.}.  Therefore, 
in such a scenario, it is important to reexamine the loop contributions 
to $Zb\bar b$ coupling and further,  the $Z$-peak constraints.  This is 
the aim of this letter. Through explicit calculations, we do find that 
gluino-sbottom loops comprising of sbottoms and a light gluino cause 
large effects on $Z$-peak observables. To account for the $R_b$ data, 
subtle cancellation between  $\tilde{b}_1$ loops and  $\tilde{b}_2$
loops is needed, which can be realized by requiring the mass splitting 
between two sbottoms  not to be too large. Numerical results show  that 
for $\tilde{b}_1$ with mass of  $2 \sim 5.5$ GeV, $\tilde b_2$ must be 
lighter than  $125$ GeV and $195$ GeV at $2\sigma$ and $3\sigma$ level, 
respectively.  
\vspace*{.2cm}
    
\noindent{\bf Calculations}~~
We start the calculations by writing down the sbottom mass-square 
matrix~\cite{susyint}
\begin{equation}
M_{\tilde b}^2=\left(\begin{array}{cc}
M_{{\tilde b}_{LL}}^2& M^{2\dagger}_{{\tilde b}_{LR}}\\
M^2_{{\tilde b}_{LR}}& M_{{\tilde b}_{RR}}
           \end{array} \right), 
\end{equation}
where $M_{{\tilde b}_{LL}}^2=M_{\tilde Q}^2+m_b^2-m_Z^2(\frac{1}{2}
-\frac{1}{3}\sin^2\theta_W)\cos(2\beta)$,  $M_{{\tilde b}_{RR}}^2= 
M_{\tilde D}^2+m_b^2-\frac{1}{3} m_Z^2 \sin^2\theta_W\cos(2\beta)$, and 
$M_{{\tilde b}_{LR}}^2=m_b(A_b-\mu\tan\beta)$.  Here $M_{\tilde Q}^2$ 
and $ M_{\tilde D}^2$ are soft-breaking mass terms for left-handed 
squark doublets $\tilde Q$ and right-handed  down squarks, respectively. 
$A_b$ is the coefficient of the trilinear term $H_1 \tilde Q \tilde D$ 
in soft-breaking terms and $\tan\beta=v_2/v_1$ is the ratio of 
the vacuum expectation values of the two Higgs doublets.
By diagonalizing the sbottom mass-square matrix, one obtains the 
physical mass eigenstates ${\tilde b}_{1, 2}$
\begin{eqnarray} 
\label{rotate}
\left (\begin{array}{l}
       \tilde b_1\\ \tilde b_2
       \end{array} \right )=\left (
             \begin{array}{cc}
            \cos\theta       &\sin\theta\\
           -\sin\theta       &\cos\theta\\
           \end{array} \right )
               \left (\begin{array}{l}
               \tilde b_L\\ \tilde b_R
              \end{array} \right ).
\end{eqnarray}
where $\theta$ is the mixing angle of sbottoms.  In  our following 
analyses we  take the sbottom masses and the mixing angel as free parameters
since they are independent of each other and determined by SUSY parameters 
$M_{\tilde{b}_{LL}}^2$, $M_{\tilde{b}_{RR}}^2$ and $M_{\tilde{b}_{LR}}^2$.

The coupling of $Z$-boson to sbottoms is given by
\begin{eqnarray}
V^{\mu}(Z\tilde{b}_i\tilde{b}^*_j)=i e O_{ij} (p_1+p_2)^{\mu},
\end{eqnarray} 
where $p_{1,2}^{\mu}$ are the momentum of $\tilde{b}_i$ and  $\tilde{b}_j$, 
respectively. $O_{ij}$ are defined as 
$O_{11}=  v_b + a_b \cos 2\theta$,
$O_{22}=  v_b - a_b \cos 2\theta$ and 
$O_{12}= O_{21}=-a_b \sin 2\theta$.
Here $v_b=1/(4 \sin\theta_W \cos\theta_W) (1-\frac{4}{3} \sin^2 \theta_W)$ and 
$a_b=1/(4 \sin \theta_W \cos \theta_W)$ are the vector and axial vector 
couplings of $Zb\bar b$, respectively.

Apparently, a light sbottom $\tilde{b}_1$ (a few GeV) can affect $Z$-peak 
observables in two ways: 
(1) the direct pair production of $\tilde{b}_1$ through $Z\tilde{b}_1\tilde{b}^*_1$ 
coupling, as discussed in  \cite{doelectr}; 
(2) the loop effects of  $\tilde{b}_1$. If gluino is also light ($12\sim 16$ GeV), 
then the loop effects are mainly from gluino-sbottom loops in  $Z b \bar{b} $ 
vertex, which comprise a light gluino $\tilde g$  and  sbottoms, as shown in Fig.~1.
It should be noted that even if the direct pair production of $\tilde{b}_1$ is 
avoided by tuning the mixing angle  $|\cos\theta| \simeq \sqrt{2/3}\sin{\theta_W} 
\simeq 0.38$ to set $Z\tilde{b}_1\tilde{b}^*_1$ coupling to be zero ($O_{11}\sim 0$),
$Z\tilde{b}_1\tilde{b}^*_2$ and $Z\tilde{b}_2\tilde{b}^*_2$ couplings still exist and
the irreducible loops shown in Fig.~1(b)  make contributions. It should also be noted that
the self-energy loops in  Fig.~1(a) involve only SUSY QCD interactions, i.e., 
gluino-sbottom-bottom couplings, which are not affected by the zero 
$Z\tilde{b}_1\tilde{b}^*_1$ coupling.        

\begin{figure}[htb]
\begin{center}
\epsfig{file=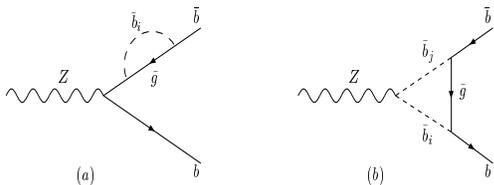,width=200pt,height=80pt}  
\caption{The gluino one-loop diagrams for $Zb\bar{b}$.}
\label{fey}
\end{center}
\end{figure}

Using dimensional regulation and adopting the on-shell renormalization scheme for 
the calculation of  Fig.~1, we obtain the effective $Zb\bar{b}$ vertex
\begin{eqnarray} \label{vertex}
V_{\mu}^{eff} (Zb\bar{b})&=&i e \left\{  
 \gamma_\mu(v_b- a_b \gamma_5)
+\frac{\alpha_s}{3\pi} \left[F_1\gamma_\mu +F_2 \gamma_\mu \gamma_5 \right. \right. 
 \nonumber\\
& & \left. \left.
+i~F_3~\sigma_{\mu\nu}k^\nu
+i~F_4~\sigma_{\mu\nu}k^\nu\gamma_5\right] \right\}.
\end{eqnarray}
Here $F_i$ are form factors originated from loop corrections, given by
\begin{eqnarray}
F_1&=&2\sum\limits_{i,j=1}^2O_{ij}\left\{
-A^-_{ij}m_bm_{\tilde{g}}\left(C_{0}(i,j)+C_{11}(i,j)\right)
\right.\nonumber\\
&&\left.+A^+_{ij}\left[m_b^2\left(C_{11}(i,j)+C_{21}(i,j)\right)
+C_{24}(i,j)\right]\right\}\nonumber\\
&&+v_b \delta Z_V+a_b\delta Z_A,\\
F_2&=&2\sum\limits_{i,j=1}^2 O_{ij} B^+_{ij}C_{24}(i,j)
-v_b \delta Z_A- a_b \delta Z_V,\\
F_3&=-&\sum\limits_{i,j=1}^2 O_{ij}\left\{
A^+_{ij}m_b\left[C_{11}(i,j)+C_{21}(i,j)\right]\right.\nonumber\\
&&\left.-m_{\tilde{g}}A^-_{ij}
\left[C_0(i,j)+C_{11}(i,j)\right]\right\},\\
F_4&=&\sum\limits_{i,j=1}^2 O_{ij} \left\{
B^+_{ij}m_b\left[2C_{12}(i,j)-C_{11}(i,j)
-C_{21}(i,j)\right.\right.\nonumber\\
&&\left.\left.+2C_{23}(i,j)\right]+m_{\tilde{g}}B^-_{ij}
\left[C_0(i,j)+C_{11}(i,j)\right]\right\} ,
\end{eqnarray}
where
\begin{eqnarray}
\delta Z_V&=&\sum\limits_{i=1}^2\left[A_{ii}^+
\left(B_1(i)+2m_b^2\frac{\partial B_1(i)}{\partial p_b^2}\right)
\right.\nonumber\\
&&\left.-2m_bm_{\tilde{g}}A_{ii}^-
\frac{\partial B_0(i)}{\partial p_b^2}\right]
\left \vert_{p_b^2=m_b^2} , \right.\\
\delta Z_A&=&-\sum\limits_{i=1}^2 B_{ii}^+B_1(i) .
\end{eqnarray}
Here $B_{0,1}(j) = B_{0,1}(-p_b, m_{\tilde{g}}, m_{\tilde{b}_j})$ and 
$C_{0,nm}(i,j)=C_{0,nm}
(-p_b, k, m_{\tilde g },  m_{\tilde b_i}, m_{\tilde b_j})$,
with  $p_b$ and $k$ denoting the four-momentum of  b quark and Z boson 
respectively, are the Feynman loop integral functions and their expressions 
can be found in \cite{bcanaly}.
Other constants appearing above are defined by   
$A_{ij}^\pm=a_ia_j\pm b_ib_j$, $B_{ij}^\pm =a_ib_j\pm a_jb_i$,
$a_{1,2}=(\sin\theta\mp\cos\theta)/\sqrt{2}$ and
$b_{1,2}=(\cos\theta\pm\sin\theta)/\sqrt{2}$.
\vspace*{.2cm}

\noindent{\bf Numerical results}~~  Let's now evaluate the effects of 
the above corrections to $Z$ peak observables. 
We start with $R_b\equiv \Gamma(Z\to b\bar b)/ \Gamma(Z\to hadrons)$.
From Eq. (\ref{vertex}) we obtain the contribution to  $R_b$
\begin{equation}
\delta R_b=R_b^{SM}(1-R_b^{SM})\Delta_{\rm SUSY}, 
\end{equation}
where 
\begin{eqnarray} 
\Delta_{\rm SUSY} &=&\frac{2\alpha_s}{3\pi}
\frac{1}{v_b^2(3-\beta^2)+2 a_b^2\beta^2}\left[v_b (3-\beta^2){\rm Re} F_1\right.
\nonumber\\
&&\left.-2 a_b \beta^2 {\rm Re} F_3+6m_b v_b {\rm Re} F_4\right]
\end{eqnarray}
with  $\beta =\sqrt{1-4m_b^2/m_Z^2}$.

To obtain  numerical results, we set input parameters as \cite{Erler:sa}
$R_b^{exp}=0.21642\pm 0.00065$, $R_b^{SM}=0.21573\pm 0.0002$,
$\sin^2\theta_W=0.2312$,  $\alpha_s(m_Z)=0.1192$,
$m_Z=91.188$ GeV and $m_b=4.75$ GeV. 
We will vary $m_{\tilde g}$ in the range $12\sim 16$ GeV and $m_{\tilde b_1}$ 
in the range  $2\sim 5.5$ GeV as was used in \cite{Berger01}.

\begin{figure}[htb]
\begin{center}
\epsfig{file=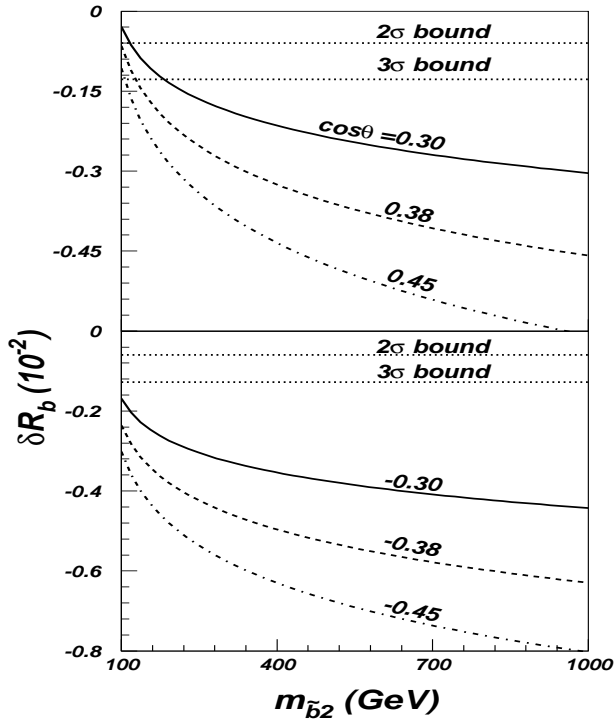,width=240pt,height=280pt}
\caption{$\delta R_b$ as a function of $m_{\tilde{b}_2} $
for $m_{{\tilde b}_1}=3.5~GeV$ and $m_{\tilde g}=14~GeV$.
The corresponding region above each horizontal line is allowed
by LEP $R_b$ data at  $2\sigma$ and  $3\sigma$ level, respectively.
}
\label{fig2}
\end{center}
\end{figure}

For $m_{{\tilde b}_1}=3.5~GeV$ and $m_{\tilde g}=14~GeV$,  
we present $\delta R_b$ versus $m_{\tilde{b}_2}$ in Fig.~\ref{fig2}.
In addition to $\cos{\theta}=\pm 0.38$ which leads to zero 
$Z\tilde b_1 \tilde b_1^*$ coupling and hence avoids large 
rate of direct pair production of $\tilde{b}_1$ \cite{doelectr}, 
we also plotted the curves for  $\cos\theta=\pm 0.30$ and $\pm 0.45$. 
From the figure, one  sees  that the contributions to  $R_b$
are negative in all the parameter space we have investigated. 
One can also see that the negative $\cos{\theta}$  gives larger 
contributions than positive one  and  as  $\vert\cos{\theta}\vert$ 
increases, the contributions become more sizable.
\vspace*{-.2cm}

Comparing  with the experimental bounds shown in Fig.~\ref{fig2},  one 
learns that even in the favorable case of positive $\cos{\theta}$, the 
contribution to $R_b $ is too large to be allowed at 3 $\sigma$ level 
if $m_{\tilde b_2} \geq 200$ GeV. Since the heavier sbottom has not been 
observed at LEPII, and it can in principle be produced in association 
with the lighter one, its mass should probably be larger than about $200$ GeV 
\footnote{ A detailed study may be needed to make this bound quantitative.}. 
So, we conclude that the scenario of one light sbottom and 
light gluino faces severe challenge.  As to the largeness of the gluino-sbottom
loop contributions, two main reasons may account for it. One is the large 
splitting between $m_{\tilde b_1} $ and $m_{\tilde b_2}$, which leads to 
a weak cancellation between $\tilde{b}_1 $ and $\tilde{b}_2$ contributions; 
the other is the lightness of sbottom $\tilde b_1$ and gluino, which induces 
large self-energy contributions. To check our understanding, we fix 
$m_{\tilde b_2}$ and $m_{\tilde g}$ but let $m_{\tilde b_1}$ approaches to 
$m_{\tilde b_2}$. Then we do find large cancellation occurs between different 
diagrams. 

\begin{figure}[htb]
\begin{center}
\epsfig{file=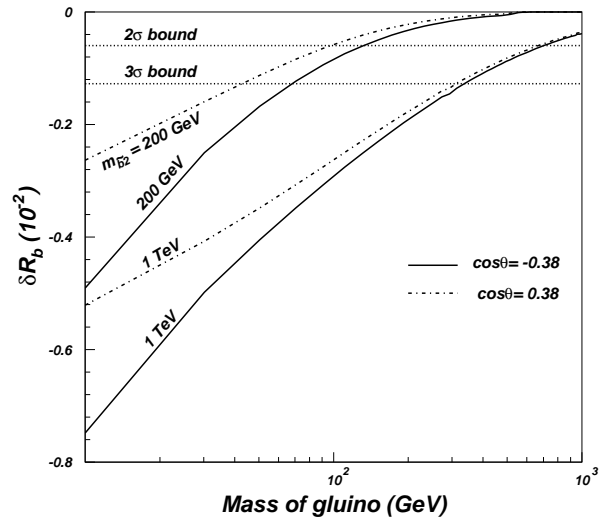,width=230pt,height=200pt}
\caption{ Same as Fig.~\ref{fig2}, but versus gluino mass for $m_{{\tilde b}_1}=3.5~GeV$.}
\label{fig3}
\end{center}
\end{figure}

We notice from Fig.~\ref{fig2}  the intriguing feature that as $m_{\tilde{b}_2}$
increases, the effects get more sizable. This can be understood as the weaker 
cancellation between $\tilde b_1 $ and $\tilde b_2$ contributions when 
$m_{\tilde{b}_2}$ increases. To further understand this behavior, we used 
approximate forms of B and C functions \cite{bcanaly}, and found that in 
the limit $m_{\tilde{b}_2}^2 \gg m_Z^2 > m_{\tilde{b}_1, \tilde{g}}^2 $, 
$\delta R_b$ is roughly linear dependent on $\ln (m_{\tilde b_2^2}
/m_{\tilde b_1^2})$ and thus increases as $m_{\tilde b_2^2}/m_{\tilde b_1^2}$ 
gets larger. Of course, this feature does not mean that SUSY QCD is 
non-decoupling from the SM.  To check the decoupling property of SUSY QCD,  
we let all relevant sparticles ($\tilde b_1$, $\tilde b_2$, $\tilde g$) become 
heavy and found the contributions drop quickly to zero.  Actually, even for 
a light $\tilde b_1$, $\delta R_b $ drops monotonously to zero when 
$m_{\tilde{g}}$ get large, as shown in Fig.~~\ref{fig3}.  

\vspace*{-.2cm}

\begin{figure}[htb]
\begin{center}
\epsfig{file=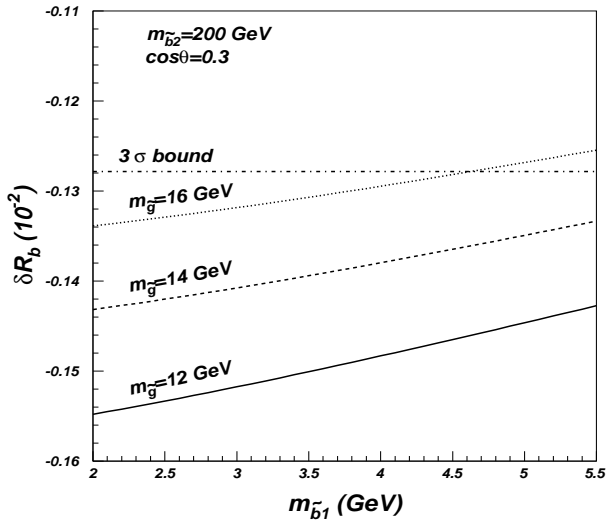,width=235pt,height=200pt}  
\caption{ Same as Fig.~\ref{fig2}, but versus  $m_{\tilde b_1}$ for 
$m_{\tilde b_2}=200$ GeV.}
\label{fig4} 
\end{center}
\end{figure}

Since in such a scenario with a light $\tilde b_1$ of a few GeV, the $\tilde b_2$ 
lighter than  $200$ GeV might be disfavored by LEP II experiment~\cite{doelectr},
we fix $m_{\tilde b_2}=200$ GeV and  $\cos{\theta}=0.3$ and plot $\delta R_b$ 
versus $m_{\tilde b_1}$ in Fig.~\ref{fig4}, where $m_{\tilde b_1}$ varies in the 
range $2\sim 5.5$ GeV and  $m_{\tilde g}$ in $12\sim 16$ GeV, as used in \cite{Berger01}  
to explain the excess of bottom quark production in hadron collision.  We see that 
such a scenario is totally excluded by the LEP $R_b$ data at $2\sigma$ level, while 
at $3\sigma$ level only a tiny corner with  $m_{\tilde g}$ close to 16 GeV and 
$m_{\tilde b_1}$ close to 5 GeV is allowed. 

Let's next consider the effects on other $Z$-peak observables:
$R_{c}$, $R_{\ell}$, $A_b$ and $A_{FB}^b$. In our calculation of these observables,  
we neglect SUSY QCD correction to $\Gamma(Z\to q \bar q)~ (q \neq b)$ since
the corresponding loops involve squarks $\tilde q$  ($\tilde q \neq \tilde b)$ 
which are assumed to be heavy.  Then the effects on all these observables 
stem only from the corrections to $Zb\bar b$ vertex in Eq.(\ref{vertex}).
Since $F_{1,2}$ are found to be much larger than $F_{3,4}$,
we neglect $F_{3,4}$ in the calculation of  $A_b$ and $A_{FB}^b$.   
In Table 1, we show the effects on these observables including $R_b$.
We see that gluino-sbottom loop effects significantly enlarge the 
deviations of the predictions from the experimental values.

\vspace*{-.4cm}
\null
\noindent
{\small Table 1: Deviation of some Z-peak observables from experimental values. 
The MSSM predictions are obtained by including SUSY QCD contributions with 
$m_{\tilde b_1}=3.5$ GeV and  $m_{\tilde g}=14$ GeV. The SM predictions are 
taken from \cite{LP01}. The values of $m_{\tilde b_1}$  are in units of GeV. }
\vspace*{0.1cm}

\begin{tabular}{|c|c|c|c|c|c|c|c|}
\hline
    & \multicolumn{6}{c|}{MSSM} &  \\  \cline{2-7} 
    &  \multicolumn{3}{c|}{$\cos\theta=0.30$} 
    &  \multicolumn{3}{c|}{$\cos\theta=0.45$} & SM\\  \cline{2-7}  
    & $m_{\tilde b_2}$: &  &  &  &  &  & \\
    & 150& 200 & 250 &  150& 200 & 250 & \\ \hline
$R_b$ &2.66$\sigma$  &3.22$\sigma$ &3.59$\sigma$ &4.49$\sigma$ 
                             &5.47$\sigma$ &6.16$\sigma$ &1.12$\sigma$ \\ \hline
$R_c$ &-0.19$\sigma$&-0.22$\sigma$  & -0.24$\sigma$ &-0.28$\sigma$ 
                           &-0.33$\sigma$ &-0.36$\sigma$ &-0.12$\sigma$\\ \hline  
$R_{\ell}$ &2.26$\sigma$&2.66$\sigma$  &2.93$\sigma$ &3.65$\sigma$ 
                           &4.32$\sigma$ &4.83$\sigma$ &1.11$\sigma$\\ \hline  
$A_b$ & -0.90$\sigma$&-0.93$\sigma$ &-0.94$\sigma$ &-0.76$\sigma$ 
                           &-0.80$\sigma$ &-0.84$\sigma$ &-0.64$\sigma$\\ \hline  
$A^b_{FB}$ & -3.25$\sigma$&-3.28$\sigma$ &-3.30$\sigma$ &-3.06$\sigma$ 
                           &-3.12$\sigma$ &-3.16$\sigma$ &-2.90$\sigma$\\ \hline 
\end{tabular}
\vspace*{0.2cm}

We should remind that in the calculation we only considered the SUSY QCD loops, i.e., 
gluino-sbottom loops. Since we focused on a special scenario of the MSSM, in 
which there exist a very light sbottom ($2\sim 5.5$ GeV) and a very light gluino 
($12\sim 16$ GeV), such gluino-sbottom loop effects are much larger than SUSY 
electroweak corrections \cite{Djouadi91}. In fact, we re-calculated SUSY EW 
corrections to $R_b$ and found they are indeed small under the current experimental 
limits on the masses of charginos and stops. The dominant contributions 
from chargino loops are found to be positive (opposite 
to SUSY QCD corrections) and at the level of $10^{-4}$,
which are about one order smaller than our present SUSY QCD 
corrections. 
\vspace*{.2cm}
    
\noindent{\bf Conclusions }~~ 
 From the above analyses we conclude that the scenario of the MSSM 
with one light sbottom ($2\sim 5.5$ GeV) and light gluino ($12\sim 16$ GeV) 
can give rise to large effects on $Zb\bar b$ vertex through gluino-sbottom
loops. Such effects significantly enlarge the deviations of some $Z$-peak 
observables, especially $R_b$,  from their experimental data. To account for 
the $R_b$ data in this scenario, the other mass eigenstate of sbottom, i.e., 
the heavier one,  must be lighter than $125$ ($195$) GeV at $2\sigma$ ($3 \sigma $) 
level, which should have been produced in association with the lighter one at LEP II
and will probobaly be within the reach of Tevatron Run 2.

\end{document}